Direct observation of spin-orbit coupling in iron-based superconductors.

S. V. Borisenko[1], D. V. Evtushinsky[1], I. Morozov[1,2], S. Wurmehl[1,3], B. Büchner[1,3], A. N. Yaresko[4], T. K. Kim[5], M. Hoesch[5], T. Wolf[6], N. D. Zhigadlo[7]

[1]*IFW-Dresden, Helmholtzstrasse 20, 01069 Dresden, Germany*

[2] *Lomonosov Moscow State University, GSP-1, Leninskie Gory, 119991 Moscow, Russian Federation*

[3]*Institute for Solid State Physics, TU Dresden, 01062 Dresden, Germany*

[4]*Max-Planck-Institute for Solid State Research, Heisenbergstrasse 1, 70569 Stuttgart, Germany*

[5] *Diamond Light Source, Harwell Campus, Didcot, OX11 0DE, United Kingdom*

[6] *Institut für Festkörperphysik, Karlsruhe Institute for Technology, Karlsruhe 76021, Germany*

[7]*Laboratory for Solid State Physics, ETH Zurich, 8093 Zurich, Switzerland*

**Spin-orbit coupling (SOC) is a fundamental interaction in solids which can induce a broad spectrum of unusual physical properties from topologically non-trivial insulating states to unconventional pairing in superconductors. In iron-based superconductors (IBS) its role has so far been considered insignificant with the models based on spin-** *or* **orbital fluctuations pairing being the most advanced in the field. Using angle-resolved photoemission spectroscopy we directly observe a sizeable spin-orbit splitting in** *all* **main families of IBS. We demonstrate that its impact on the low-energy electronic structure and details of the Fermi surface topology is much stronger than that of possible nematic ordering. Intriguingly, the largest pairing gap is always supported exactly by SOC-induced Fermi surfaces.**

In the presence of spin-orbit coupling, the electron's spin quantized along any fixed axis is no longer a good quantum number, but its total angular momentum is. This basic fact alone or in combination with a particular symmetry breaking may lead to a splitting of otherwise degenerate energy bands and is the origin of fascinating phenomena such as spin Hall effects [1], spin relaxation [2], topological insulation [3], Majorana fermions [4] etc. No wonder that the systems with SOC are in the focus of intensive research in the field of spintronics [5] – there is a unique opportunity to manipulate the spin without the aid of magnetic field.

A special role has been played by SOC in the field of superconductors. In low-dimensional or noncentrosymmetric systems it can promote and stabilize superconductivity [6], allow ferromagnetism to coexist with superconductivity [7] or even rise $T_c$ [8]. If SOC is large, some superconductors can host an elusive Fulde-Ferrell-Larkin-Ovchinnikov state [9] or topological superconductivity [4]. It is anticipated that SOC could be a very important ingredient in describing the superconducting state in $Sr_2RuO_4$ [10]. Since k-dependent spin-orbit splitting is larger than the superconducting gap in this material, the SOC-induced spin anisotropy together with the orbital mixing should directly influence the orbital and spin angular momentum of the Cooper pairs. Singlet and triplet states could be strongly mixed, blurring the distinction between spin-singlet and spin-triplet pairing [11].

In multiband iron-based superconductors, where the low energy electronic structure is composed of different orbitals, the situation is even more complicated because of the presence of the sizeable Hund's coupling. When the electronic structure near the Fermi energy is composed of different orbitals and spins mixed via spin-orbit coupling, determination of the pairing symmetry becomes non-trivial. However, up to now SOC in iron pnictides and chalcogenides was considered weak.

We start with the example of LiFeAs, which is a special representative of iron-based family of superconductors [12]. This material is one of the most studied due to its stoichiometry and non-polar surfaces. Its electronic structure is believed to be well understood from numerous angle-resolved photoemission experiments (ARPES) and the parameterization of its electronic dispersions has been used to test the most developed theoretical approaches [13-15]. To detect spin-orbit coupling in LiFeAs experimentally we first need to identify the region of the momentum space where this effect is predicted to be the strongest. The results of band structure calculations with and without inclusion of SOC to the computational scheme are shown in Fig.1.

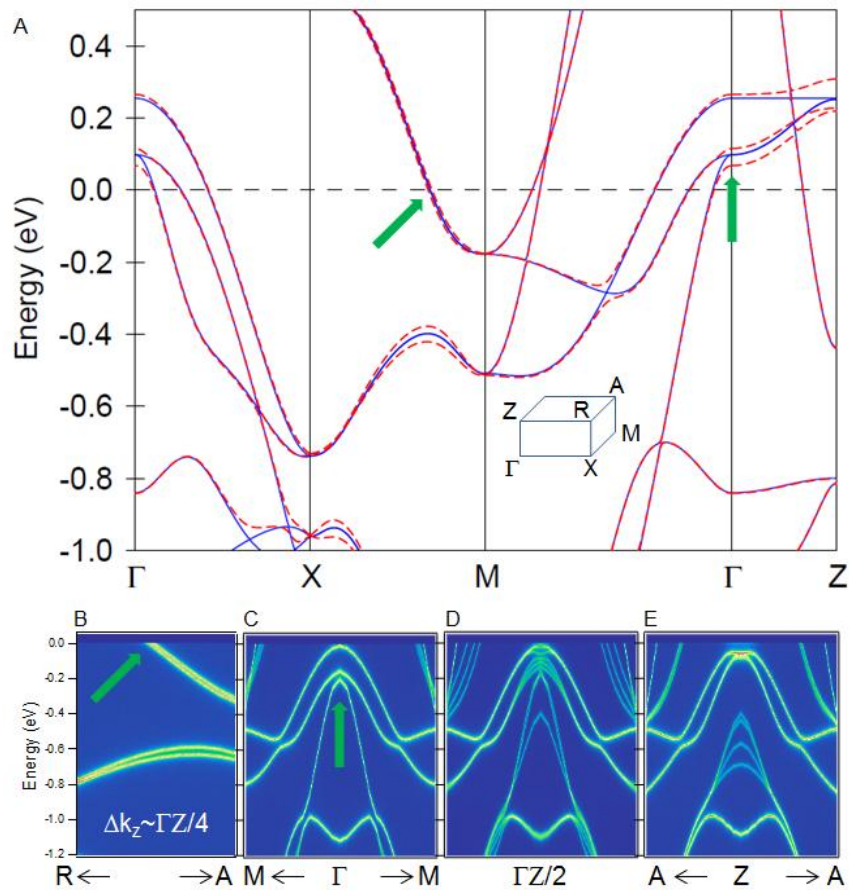

**Fig. 1. Band structure calculations of LiFeAs with and without SOC.** (A) Electronic bands of LiFeAs excluding SOC (blue lines) and including SOC (red dashed lines) along high-symmetry directions. The corresponding Brillouin zone (BZ) with high-symmetry points is shown in inset. (B) Calculations with SOC along AR direction summed along $k_z$ within $\Delta k=\Gamma Z/4$ momentum interval in false color scale to simulate ARPES signal. (C) The same as (B), along M-Γ-M direction. (D) The same as (B), along AM/2-ΓZ/2-AM/2 direction. (E) The same as (B), along A-Z-A direction. Hereafter the Fermi level of the calculations is shifted arbitrarily either to fit experiment or to reveal the behavior of the particular states.

There are several places where the effect of SOC on the band structure is large. We will not consider the splitting below 200 meV because its detection is hindered by the scattering caused by the correlations [16] even at very low temperatures. We have though detected noticeably broader energy distribution curves (EDC) in these regions which would indicate the lifting of the degeneracy [17]. The best candidates situated close to the Fermi level, where the scattering is low, are thus the states near the Γ-point and near crossing along MX direction indicated by the green arrows. In both cases SOC qualitatively changes the band structure by splitting the bands which would be degenerate if SOC is absent. Earlier it has been shown, that experimental results differ from theory in several aspects. The most pronounced of them are the strong renormalization and different Fermi surfaces without nesting [12], the latter being caused by the orbital-dependent renormalization. One important consequence is that the maxima of the *xz/yz* bands become occupied for certain $k_z$ values near the Γ-point, providing a convenient opportunity to determine the size of the SOC in LiFeAs.

ARPES momentum resolution along the axis perpendicular to the surface is significantly lower than the in-plane one because of the finite escape depth of photoelectrons, and is thus material dependent. The $k_z$-dispersion in iron-based superconductors can be relatively well defined and corresponding momentum resolution is given by a certain fraction of the BZ size along $k_z$ [18]. We show in Fig. 1, B to E the calculated data summed over BZ/8 interval along $k_z$ (discrete mesh) for several cuts through the BZ to simulate, in a first approximation, the photoemission data. We found that the most suitable $k_z$ values for determining SOC are not those corresponding to MX direction (Fig. 1A) but along AR direction (Fig. 1B) where the splitting is maximal and remains not completely blurred by $k_z$ integration. The magnitude of SOC here is expected to be smaller in comparison with the splitting in Γ-point, but still can be detected. The integrated dispersions near the center of the BZ can be grouped in three types according to their qualitative behavior (Fig. 1, C to E). Near the Γ-point all three bands are well separated and summation over $k_z$ does not result in significant broadening of the features. Energy positions of the band edges are the lowest here. Upon departing from Γ-point *xz/yz* bands are shifted up in energy. Finally, for the $k_z$'s closer to Z-point we expect rather blurred spectral weight, as the top of the *xz/yz* bands is crossed by a *$3z^2$-$r^2$* band which has a large $k_z$-dispersion and should appear in the spectra rather broadened. We note that this picture is qualitatively similar in all IBS with differences caused by value of the SOC, details of the dispersion of *$3z^2$-$r^2$* band and position of the Fermi level (see Fig. S5).

Figure 2 represents experimental data taken aiming at measurements of SOC at Γ-point. In order to detect the possible splitting under the most suitable conditions, we have to locate Γ-point in all directions of momentum. If in-plane location is defined by the normal emission, $k_z$=0 can be found by recording the $k_x$=$k_y$=0 spectrum as a function of photon energy. Such data taken in the interval 20-110 eV with the step of 0.5 eV are shown in Fig.2A. In panel (B) of the same figure we show the results of the relativistic band-structure calculations (Fig. 1A) along the ΓZ direction using the representation style of Fig. 1,B and E. Taking into account the well-known renormalization factor of 3 and square root dependence of $k_z$ from hν, the agreement is remarkable and allows us to unambiguously locate Γ- and Z-points in terms of photon energy. We note that the strongly dispersing *$3z^2$-$r^2$* band with the minimum in Z-point is seen so clearly for the first time in iron-based superconductors, but we postpone the discussion of the implications until the end of this paper. The panels (C) and (D) of Fig. 2 represent the photoemission intensity along the MΓM and AZA directions, respectively. The qualitatively different dispersions directly agree with the corresponding patterns of expected intensity distributions from Figs. 1, C and E. To clarify the picture in the immediate vicinity of the Fermi level, we recorded the spectra from the same k-space locations using the light of different polarisations and at low temperature (Fig. 2F). It is obvious, that both bands are located below the Fermi level at the Γ-point. While such two peaks at zero momentum (k=0) are clearly absent in Fig. 2D and in right panels of Fig. 2F, one weaker maximum close to the Fermi level is still present. We found that the spectral function indeed reaches its maximum *above* the Fermi level and this weaker single peak is caused by the Fermi function above $T_c$ (Fig. 2D) and by the presence of the gap below $T_c$ (right panels of Fig. 2F, see SM).

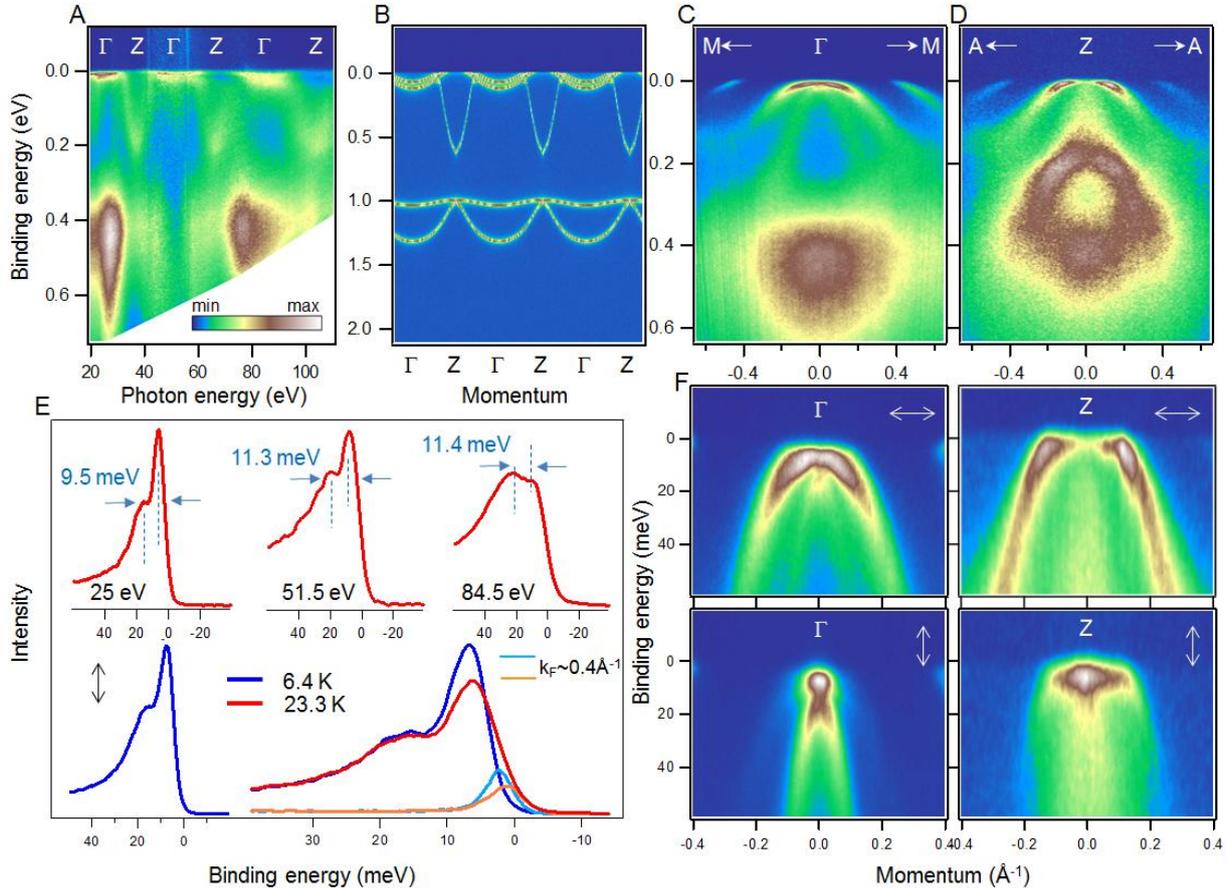

**Fig. 2. Determination of spin-orbit splitting at Γ-point.** (A) Photon-energy dependence of the normal emission EDC representing the band structure along ΓZ-direction at T = 23.3 K (B) Calculations with SOC along the ΓZ-direction for comparison with (A). (C) Energy-momentum intensity distribution recorded along the M-Γ-M direction at T = 23.3 K using the 25 eV photons. (D) Same as (C), but along A-Z-A directions using 35.5 eV photons. (E) EDCs from the Γ-point taken at T = 23.3 K (red curves) and at T = 6.4 K (blue curves) using different photon energies and polarizations. Orange and light blue $k_F$-EDCs taken above and below $T_c$ respectively are shown for comparison. (F) High-resolution low-temperature ARPES data near Γ- and Z-points recorded with the light of horizontal (upper) and vertical (lower) polarizations.

The value of SOC can now be directly read from the distance between the peaks of the EDCs, corresponding to Γ points, presented in upper panels of Fig. 2e. The average value is ~10.7 meV. We also show such EDCs recorded in the superconducting state and using the light of different polarization (lower panels of Fig. 2E) together with the single-peaked EDCs from the $k_F$~0.4 Å$^{-1}$ which correspond to the dispersions definitely crossing Fermi level (Fig. 2 C, D). It is seen that the superconducting gap does not significantly influence the splitting by slightly shifting the lower binding energy feature, which is not surprising since the maximum gap in LiFeAs is comparable with its energy position [19]. As expected, the opening of the gap has more impact on the $k_F$-EDCs, although the gap magnitude for them is lower [19].

Now we turn to the other location in the k-space mentioned earlier to measure the magnitude of the SOC on the electron pockets. One has to take into account that the photon energy will be slightly different from the one corresponding to the Z-point, since the momentum in question (crossing along AR line, see Fig. 1) is probed by slightly higher energies. We found that 21 eV is very suitable for this purpose. Since in this case the SOC split dispersions cross the Fermi level in ($k_x$, $k_y$) plane, we can observe this splitting in the momentum space with high resolution. For this purpose we recorded the detailed Fermi surface map near the corner of the BZ. Indeed, as follows from Fig.3A, electron pockets are no longer degenerate along the AR direction, contrary to what one would expect from non-relativistic band structure. To quantify the effect in terms of energy, we integrate the intensity from the same dataset in the narrow momentum region indicated by the white double-headed arrow in Fig. 3A and plot it as a function of momentum along AR and energy in panel (C). In close agreement with Fig. 1B, the dispersions are split in the experimental data shown in Fig. 3B. EDC corresponding to the white arrow is separately shown in Fig. 3D and clearly demonstrates the presence of two peaks distanced by ~10 meV. This value is about 2.4 times lower than the one predicted by the calculations, but remembering that all the bands in the vicinity of the Fermi level are renormalized in LiFeAs by a factor of 3, the energy splitting observed in two different places of BZ is very reasonable. Again, we would like to be certain that the effect is not temperature dependent and compare the present result with the data recorded deep in the superconducting state (Fig. 3C). In spite of the presence of the superconducting gap, the momentum distribution of the intensity below $T_c$ is qualitatively the same, also demonstrating the lifting of the degeneracy of electron pockets along the high symmetry directions.

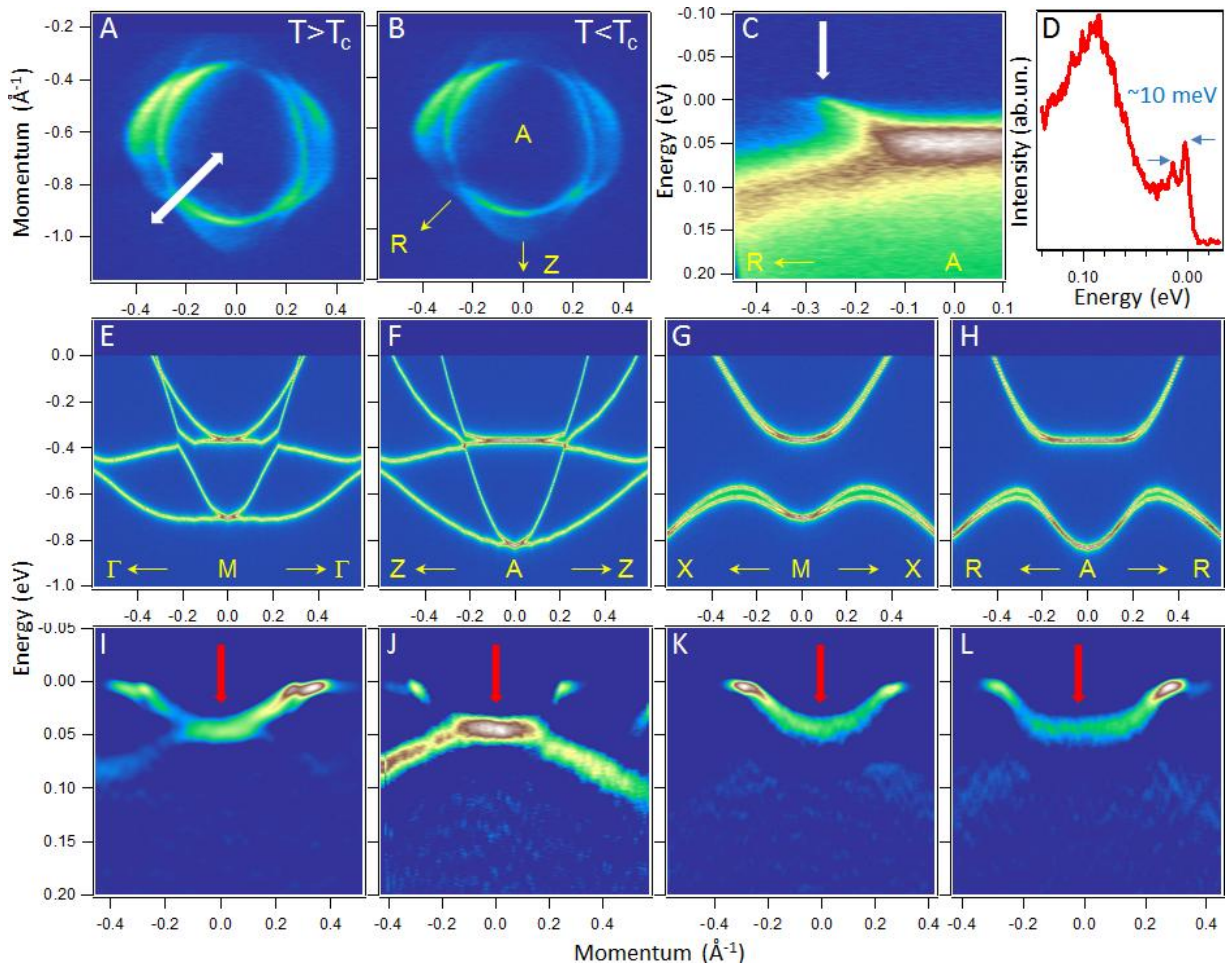

**Fig. 3. Spin-orbit splitting on electron-like pockets and absence of nematic splitting.** (A) High-precision Fermi surface map of electron pockets in LiFeAs recorded at 23 K (T>$T_c$). White arrows show the direction to high-symmetry points. Double-headed arrow indicates the cut in momentum space along which the SOC is the most noticeable. (B) Momentum-energy cut corresponding to the double-headed arrow in (A). Thick arrow shows the momentum at which the splitting is the most obvious. (C) The same as (A), but recorded at 6.4 K (T<$T_c$). (D) EDC corresponding to the white thick arrow in (B). (E) to (H) Results of the band structure calculations along the cuts passing through M and A points. (I) to (L) Corresponding experimental data represented by the second derivative along energy axis of the raw data. The latter are shown in Fig. S6. Red arrows indicate the single feature and absence of any splitting.

Before switching to other IBS families, we have to exclude an alternative reason for the splitting at the center of the BZ, which has been suggested recently [20], based on the developing of nematic ordering at low temperatures. We do it with the help of theoretical study [21] where the influence of both SOC and nematic ordering on the electronic structure is considered. According to this study, while the splitting at Γ-point can be a result of both effects, there should be a splitting of each of the bottoms of the electron pockets at M-point if nematicity is present, while degeneracy of both bottoms is protected if only SOC is present (Fig. 1 A). For this purpose we have recorded the spectra along different cuts crossing the MA line and compared them with the calculations (Fig. 3 E to L). While the bottom of the deeper pocket is very broadened by scattering and cannot be used to address the question, one is still able to consider the shallower one. As follows from Fig. 3 E to L there is no detectable splitting at these points. Moreover, the width of the corresponding EDC (Fig. S7 C) is typical for this binding energy, i.e. it is defined by many-body effects, thus implying the absence of any noticeable influence of nematic order. In addition, contributions from two domains would imply doubling of all the dispersing features (Fig. S8), which is not seen experimentally. Doping the LiFeAs with Co, as in all other IBS, results in broadening of the features because of the additional scattering on impurities and detection of the SOC in those materials is therefore more difficult. For the sake of comparison with the results of Ref. 20, we have recorded the EDC in Γ-point of 10%-doped compound and found comparable value of SOC. As mentioned above, the temperature does not influence the observed splitting in LiFeAs. However, further increase of temperature would obviously hinder the observation of ~10 meV splitting of two features with intrinsic width of ~5 meV each, which will merge into a seemingly single broad peak [20].

The decisive evidence to support the dominant role of spin-orbit interaction in comparison with the nematic effects comes from the similar experiments on FeSe, the simplest IBS. In Fig. 4 A to F we show the spectra taken along the cuts passing through the Γ, Z and M points and compare them with the relativistic band-structure calculations (see also Fig. S5 A). There is a close qualitative correspondence between the theory and experiment as far as the behavior of the dispersing features is concerned, as in LiFeAs. The peculiar $k_z$ dispersion of the states near the Fermi level is strongly influenced by the presence of the $3z^2-r^2$ band and such an agreement with the experiment clearly implies its presence also in FeSe. The SOC in FeSe is predicted to be larger than in LiFeAs and this is immediately seen experimentally (~25 meV in FeSe vs 11 meV in LiFeAs). If this larger splitting was due to the orbital ordering effects, one would expect it to be clearly seen in M-point. As Fig. 4 F shows, in FeSe one is able to track the behavior of both shallow and deep electron pockets because both are located at lower than in LiFeAs binding energies (lower scattering). Again, in both cases the bottoms of the electron pockets remain degenerate. Moreover, now we can determine even more precisely the upper limit of the possible nematic splitting. The corresponding EDC ( Fig. S7 B) has two peaks separated by ~ 10 meV. Taking into account the intrinsic widths of these peaks the possible remaining splitting of each of them, not detected because of finite energy resolution, is of the order of 2-3 meV. In addition, as in the case of LiFeAs, we do not see any doubling of the dispersion features because of the possible domains [21].

We also present the analogous data for other two main families of IBS. In Fig. 4 G the spin-orbit splitting is clearly seen in overdoped Co-BaFe$_2$As$_2$ (T$_c$~10 K) where it is possible to see both (originally *xz* and *yz*) bands below the Fermi level at $\Gamma$-point. In case of the optimally hole-doped 122 material (T$_c$~38 K), there is no possibility to determine the SOC directly since tops of all hole-like bands are well above the Fermi level in accordance with lower electron concentration. We go around this limitation by the following way. As was found out earlier [22], because of the sizeable superconducting gap and proximity of the band's edges to E$_F$, the top of the band is "reflected" to the occupied side of the spectrum below T$_c$, as schematically shown in Fig. 4H. The SOC can be then determined from the corresponding EDC (Fig. S7 D). Another peculiarity of the 122 family is that the SOC now lifts the degeneracy in the corner of the BZ (now X-point, not M, since BZ is different) because of different crystal structure. This splitting is visible in Fig. 4 I. Finally, we detect the splitting in a representative of 1111 family, Co-SmFeAsO (T$_c$~16K). In accordance with the calculations, there is a doublet in $\Gamma$ and a singlet in M-point (Fig. 4 J and K). We summarize our observations in Fig. 4 L where we plot the experimentally determined values together with those predicted by the band structure calculations. There is a clear correlation between two datasets which speaks in favor of correctness of our interpretation.

Detected SOC and presence of the *3z$^2$-r$^2$* band at the Fermi level drastically change our knowledge of the low energy electronic structure and Fermi surface in IBS. In all cases we observed the decisive influence of the spin-orbit interaction on the low-energy electron dynamics. While in 11, 111 and electron-doped 122 families the singularities near center of the BZ are SOC-induced, in hole-doped 122, at a first glance, it is the complicated structure near the corner of the BZ which is strongly influenced by SOC. If one recalls (Fig. 4 H) that the tops of the SOC split bands in hole-doped 122 are very close to E$_F$ from unoccupied part of the spectrum, it becomes clear that also electrons near the center of the BZ are better described by considering SOC. This is even more obvious in 1111 family where both the tiny hole and electron pockets are formed by SOC split states (Fig. S5 C). Fermi surface in all IBS becomes very sensitive to doping level and essentially three-dimensional taking into account the manifold of bands and their interaction along $\Gamma$Z in all systems.

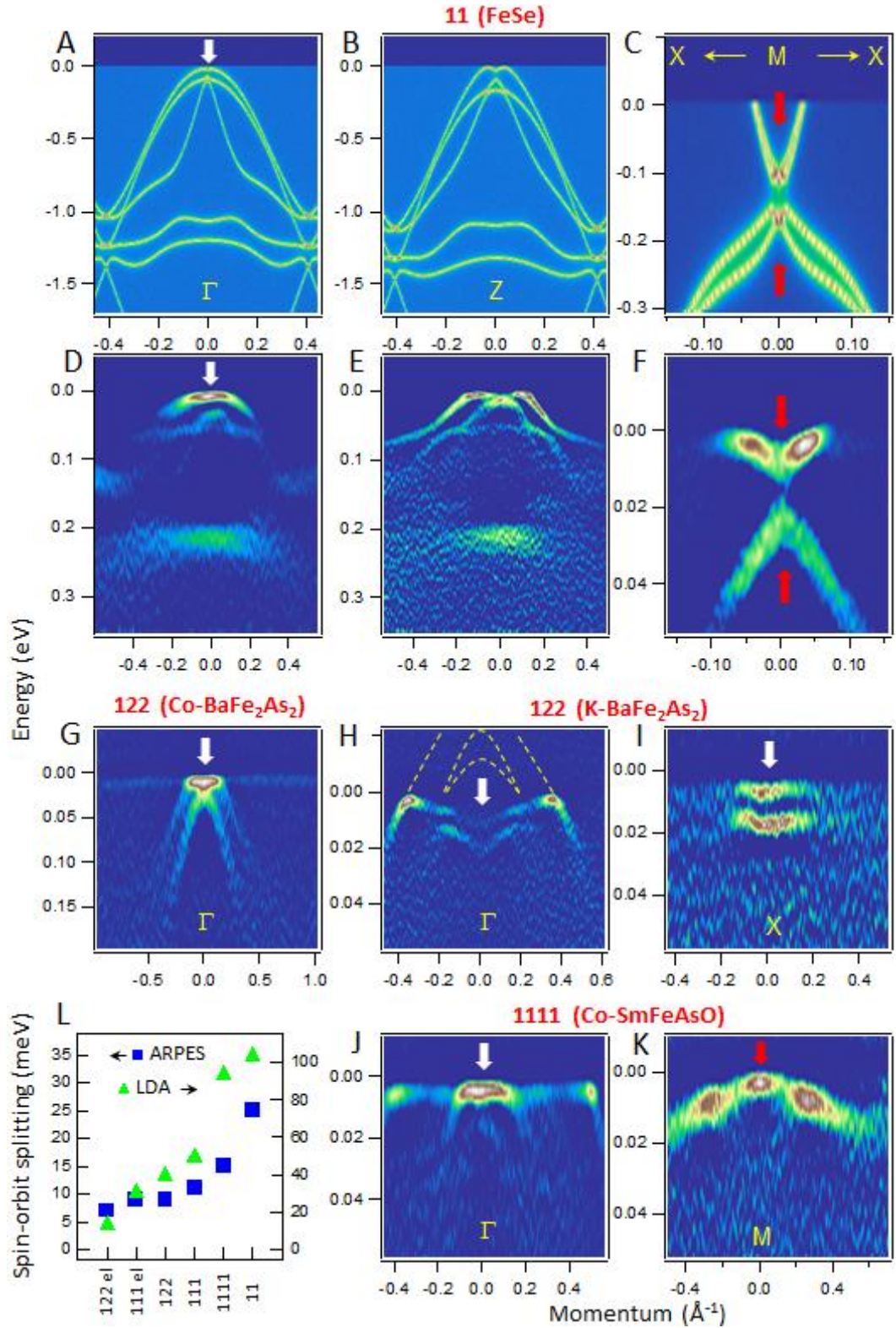

**Fig. 4. Spin-orbit coupling in other iron-based superconductors.** (A) to (C) Results of the relativistic band structure calculations of FeSe along the high-symmetry directions. (D) to (F) Corresponding experimental data, shown as second derivatives of the raw data. Note, that in panel (C) two singlets are observed, contrary to two doublets expected in nematic scenario. (G) to (I) The same for 122 materials. Dashed lines in panel (H) show the anticipated dispersions in the unoccupied part of the spectrum. (J) and (K) The same for Co-SmFeAsO. (L) Comparison of the experimental values for SOC obtained by reading the peak positions from corresponding to panels (D) to (K) EDCs shown in Fig. S7, with the theoretical values. "el" means electron pocket. White arrows are intended to illustrate that there is a doublet, while red arrows show single features. Corresponding raw data are presented in Fig. S9.

The present result sheds light on the several earlier experiments. NMR and STM studies of LiFeAs reported the behavior of the Knight shift and quasiparticle interference, respectively, consistent with the triplet pairing for particular samples [23, 24]. Indeed, it has been shown that combined effect of Hund's interaction and SOC can lead to spin-triplet superconductivity in multi-orbital systems [11]. Another theoretical study demonstrates that inclusion of SOC not only leads to additional mixing of triplet pairing, but also to anisotropic energy gap on all Fermi surfaces in iron-based superconductors [25], the latter being observed experimentally [19]. SOC causes a momentum-dependent disparity between in-plane and out-of-plane components of the spin susceptibility. This disparity is maximal at SDW wave vector and is present in both normal and SC states [26]. In the former the in-plane components win, which is consistent with the observation of in-plane orientations of spins in the magnetic SDW state. SOC also provides a qualitative explanation of the spin-space asymmetry in neutron scattering and NMR data [27-31]. Spin-resonance for s± gap symmetry is partially suppressed in the out of-plane component as compared to the in-plane one [32].

Our results should also be taken into account in many theoretical models which consider the orbital ordering as relevant to the physics of iron-based superconductors. Presence of SOC does not allow classifying electronic states relevant for superconductivity in accordance with the atomic orbitals, especially near $\Gamma$-point, as is illustrated in Fig. S10. If SOC is absent, *xz* and *yz* orbitals are degenerate. SOC lifts this degeneracy and mixes them into *xz + iyz*, and *xz – iyz*. Upon moving away from $\Gamma$ the influence of SOC is decreasing.

Although the new results call for more precise measurements of the gap function of IBS, it is obvious that in all known to us cases it is the SOC-induced Fermi surfaces which support the largest pairing gap, as e.g. Z-pocket in LiFeAs [19]. We hope that the refined electronic structure and gap function in iron-based materials will bring the comparison with the theoretical models to a new level [13-15]. Detected SOC in LiFeAs is a general property of all iron pnictides and chalcogenides. Recalling that all of these materials with considerable critical temperatures are characterized by the singular Fermi surfaces [33-36], thus implying that the Fermi energy is of the order of gap and SOC, the latter may have profound implications on the mechanism of superconductivity defining the detailed shape and orbital character of the Fermi surface sheets.

## Acknowledgements

We thank Maxim Dzero, George Jackeli, Vladimir Antropov, Hajo Grafe and Markus Braden for helpful discussions and to Robert Beck for performing magnetization measurements. The work was supported under grants No. BO1912/2-2, BO1912/3-1, BE1749/13, WU595/3-1. I.M. is grateful for support through RFBR grant No 15-03-99628a.

## References

[1] T. Jungwirth, J. Wunderlich & K. Olejnik, *Spin Hall effect devices.* Nat. Mater. 11, 382 (2012)

[2] B. Zimmermann et al. *Anisotropy of Spin Relaxation in Metals.* Phys. Rev. Lett. 109, 236603 (2012)

[3] M. Z Hasan, C. L Kane: *Topological Insulators*, Rev.Mod.Phys. 82, 3045 (2010)

[4] V. Mourik et al. *Signatures of Majorana Fermions in Hybrid Superconductor-Semiconductor Nanowire Devices.* Science 336, 1003 (2012)

[5] S. A. Wolf et al. *Spintronics: A Spin-Based Electronics Vision for the Future.* Science 294, 1488 (2001)

[6] A. D. Caviglia et al. *Tunable Rashba Spin-Orbit Interaction at Oxide Interfaces.* Phys. Rev. Lett. 104, 126803 (2010)

[7] D. A. Dikin et al. *Coexistence of Superconductivity and Ferromagnetism in Two Dimensions.* Phys. Rev. Lett. 107, 056802 (2011)

[8] H. Jeffrey Gardner et al. *Enhancement of superconductivity by a parallel magnetic field in two-dimensional superconductors.* Nat. Phys. 7, 895–900 (2011)

[9] D. F. Agterberg *Novel magnetic field effects in unconventional superconductors.* Physica C 387, 13 (2003)

[10] M. W. Haverkort et al. *Strong Spin-Orbit Coupling Effects on the Fermi Surface of Sr2RuO4 and Sr2RhO4.* Phys. Rev. Lett. 101, 026406 (2008)

[11] C. M. Puetter and H.-Y. Kee *Identifying spin-triplet pairing in spin-orbit coupled multi-band superconductors.* EPL 98, 27010 (2012)

[12] S. V. Borisenko et al. *Superconductivity without Nesting in LiFeAs.* Phys. Rev. Lett. 105 (6), 067002 (2010)

[13] Y. Wang et al. *Superconducting gap in LiFeAs from three-dimensional spin-fluctuation pairing calculations.* Physical Review B 88 (17), 174516 (2013)

[14] F. Ahn et al. *Superconductivity from repulsion in LiFeAs: Novel s-wave symmetry and potential time-reversal symmetry breaking.* Physical Review B 89 (14), 144513 (2014)

[15] T. Saito et al. *Reproduction of experimental gap structure in LiFeAs based on orbital-spin fluctuation theory: s++-wave, s±-wave, and hole-s±-wave states.* Phys. Rev. B 90, 035104 (2014)

[16] D. V. Evtushinsky et al. *Anomalous High-Energy Electronic Interaction in Iron-Based Superconductor.* arXiv:1409.1537

[17] Supplementary information

[18] D. V. Evtushinsky et al. *Strong electron pairing at the iron 3dxz,yz orbitals in hole-doped BaFe2As2 superconductors revealed by angle-resolved photoemission spectroscopy.* Phys. Rev. B 89, 064514 (2014)

[19] S. V. Borisenko et al. *One-Sign Order Parameter in Iron Based Superconductor.* Symmetry 4, 251-264 (2012)

[20] H. Miao et al. *Coexistence of orbital degeneracy lifting and superconductivity in iron-based superconductors.* Phys. Rev. B 89, 220503(R) (2014)

[21] Rafael M. Fernandes and Oskar Vafek. *Distinguishing spin-orbit coupling and nematic order in the electronic spectrum of iron-based superconductors.* Phys. Rev. B 90, 214514 (2014)

[22] D. V. Evtushinsky et al. *Fusion of bogoliubons in $Ba_{1-x}K_xFe_2As_2$ and similarity of energy scales in high temperature superconductors*. arXiv:1106.4584

[23] S.-H. Baek et al. *75As NMR-NQR study in superconducting LiFeAs.* Eur. Phys. J. B 85, 159 (2012)

[24] T. Hänke et al. *Probing the Unconventional Superconducting State of LiFeAs by Quasiparticle Interference.* Phys. Rev. Lett. 108, 127001 (2012)

[25] V. Cvetkovic and O. Vafek *Space group symmetry, spin-orbit coupling, and the low-energy effective Hamiltonian for iron-based superconductors.* Phys. Rev. B 88, 134510 (2013)

[26] M. M. Korshunov, Y. N. Togushova, I. Eremin, P. J. Hirschfeld *Spin-Orbit Coupling in Fe-Based Superconductors.* Journal of Superconductivity and Novel Magnetism 26, 2873 (2013)

[27] N. Qureshi et al. *Local magnetic anisotropy in BaFe2As2: A polarized inelastic neutron scattering study.* Phys. Rev. B 86, 060410 (2012)

[28] P. Steffens et al. *Splitting of Resonance Excitations in Nearly Optimally Doped Ba(Fe0.94Co0.06)2As2: An Inelastic Neutron Scattering Study with Polarization Analysis.* Phys. Rev. Lett. 110, 137001 (2013)

[29] F. Wasser et al. *Spin reorientation in Na-doped BaFe2As2 studied by neutron diffraction.* arXiv:1407.1417

[30] O. J. Lipscombe et al. *Anisotropic neutron spin resonance in superconducting BaFe1.9Ni0.1As2.* PRB 82, 064515 (2010)

[31] K. Matano et al. *Anisotropic spin fluctuations and multiple superconducting gaps in hole-doped Ba0.72K0.28Fe2As2: NMR in a single crystal.* EPL 87, 27012 (2009)

[32] M. Liu et al. *Polarized neutron scattering studies of magnetic excitations in electron-overdoped superconducting BaFe1.85Ni0.15As.* PRB 85, 214516 (2012)

[33] J. Maletz et al. *Photoemission and muon spin relaxation spectroscopy of the iron-based Rb0.77Fe1.61Se2 superconductor: Crucial role of the cigar-shaped Fermi surface.* Phys. Rev. B 88, 134501 (2013)

[34] S. Thirupathaiah, T. Stürzer, V. B. Zabolotnyy, D. Johrendt, B. Büchner, and S. V. Borisenko *Why Tc of (CaFeAs)10Pt3.58As8 is twice as high as (CaFe0.95Pt0.05As)10Pt3As8.* Phys. Rev. B 88, 140505 (2013)

[35] J. Maletz et al. *Unusual band renormalization in the simplest iron-based superconductor FeSe1−x.* Phys. Rev. B 89, 220506(R) (2014)

[36] A. Charnukha et al. preprint

**Supplementary information**

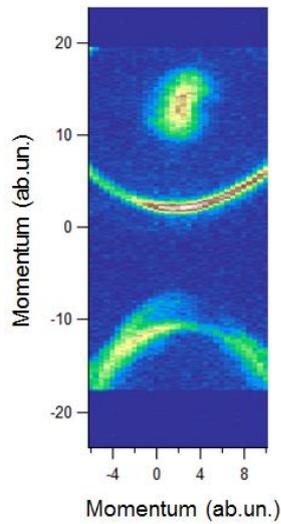

**Fig. S 1.** Fermi surface map taken at 21 eV photon energy to locate the normal emission.

In this section we present the experimental and computational details as well as several datasets supporting the conclusions drawn in the main text.

ARPES measurements were performed at the I05 beamline of Diamond Light Source, UK. Single crystal samples were cleaved in-situ at a pressure lower than 2*10-10 mbar and measured at temperatures ranging from *6.4-23.3 K*. Measurements were performed using (*s,p*)-polarised synchrotron light from 18-*120 eV* and employing Scienta R4000 hemispherical electron energy analyser with an angular resolution of *0.2-0.5 deg* and an energy resolution of *3 - 20 meV*.

Band structure calculations were performed for the experimental crystal structures (e.g. [1] for LiFeAs) in the local density approximation (LSDA) using the linear muffin-tin orbital (LMTO) method. Some details of the implementation of the PY LMTO code [2] can be found in Ref. 3. Spin-orbit coupling was added to the LMTO Hamiltonian at the variational step. Since LiFeAs does not show magnetic order ARPES spectra are compared to non-spin-polarized band structure, although LSDA calculations give a magnetic ground state with strip-like antiferromagnetic order and somewhat lower total energy.

LiFeAs single crystals in the form of packets of plates with dimensions of up to 1 cm were grown by self-flux by the standard method [4]. For the ARPES study single-crystal plate with dimensions of 3 * 3 * 0.1 mm$^3$ been selected.

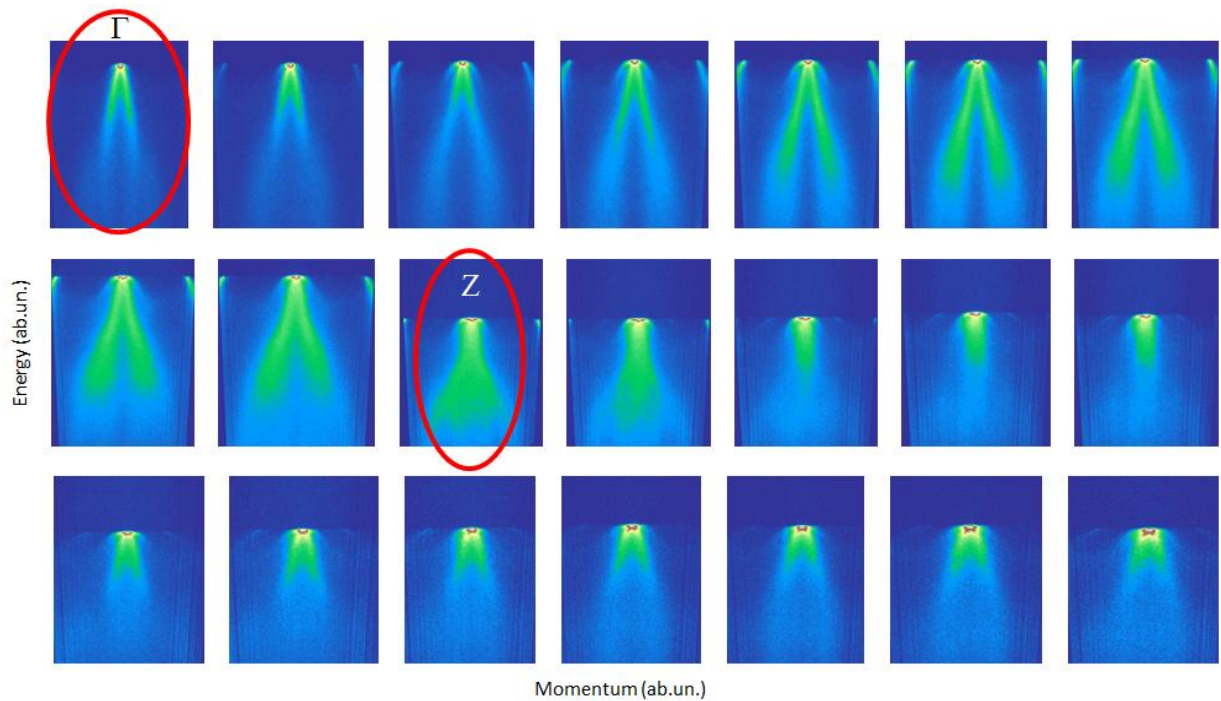

**Fig. S 2.** Exemplary datasets taken using photon energies from 85 eV to 30 eV and vertical polarisation.

Since the dispersions near $\Gamma$-point are steep in $k_x/k_y$ plane, it is crucial to determine the momentum with high precision. Any deviation would result in a splitting because of the misalignment as the xz/yz-bands are non-degenerate away from Γ-point even if the SOC is absent. To fulfill this requirement, we recorded Fermi surface map similar to the one shown in Fig. S4A, but at lower photon energy (21 eV) to ensure more precision when locating the high symmetry points. One of such maps is shown in Fig. S1. It allowed us to find the polar angle corresponding exactly to Γ-point (2.27°). Another reason to do mapping is to define the orientation of the crystal with respect to the vector of linear polarization of the beam. Earlier measurements demonstrated that the most favorable conditions to measure tops of the xz/yz bands are achieved when polarization is perpendicular to the reaction plane. We have taken the data exactly in this geometry (Fig. 2). In Fig. S2 we show exemplary datasets taken with different excitation photon energies to locate Γ and Z-points using vertical polarisation. Even without plotting the detailed photon energy dependence as in Fig. 2A, one is able to find qualitatively different intensity distributions which correspond to two high-symmetry points in question.

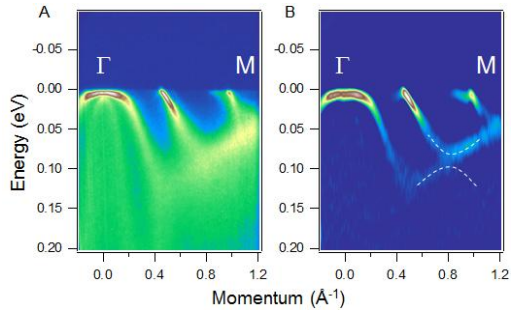

**Fig. S 3.** Raw ARPES data along the ΓM direction (left). Second derivative with respect to momentum (right). Dashed lines are guide to eye.

Fig. 1A implies that there are places at higher binding energies where the spin-orbit splitting is significant. In spite of the usual broadening because of the increased scattering, one is able to see the splitting also here. The raw data together with the second derivative plot are shown in Fig. S3.

In order to confirm that the top of the xz/yz dispersion is below the Fermi level in Γ-point and above it in Z-point, we have multiplied the spectra taken at 23.2 K by the Fermi function (Fig. S4). This procedure enhances the intensity above the Fermi level. It is obvious, that the spectral function has its maximum below (in Γ) and above (in Z) Fermi level.

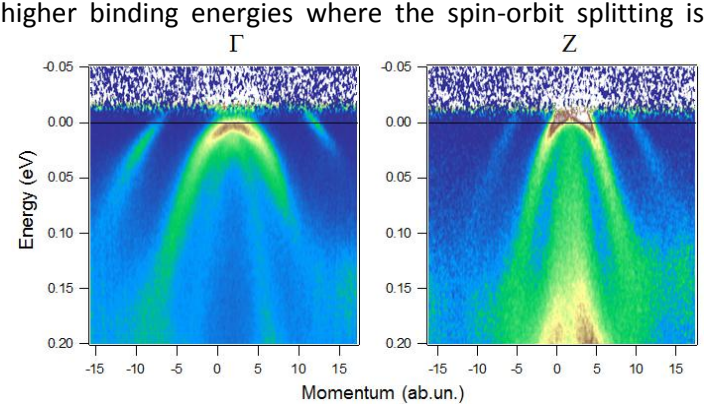

**Fig. S 4.** ARPES data recorded using 25 (left) and 36.5 eV (right) photon energy multiplied by the Fermi function.

In Fig. S5 we show the results of LDA band structure calculations for 11, 122 and 1111 parent compounds. Comparing these calculations with the ones shown in Fig.1 for 111, one can immediately notice the qualitative similarities along the ΓZ direction as far as SOC and $3z^2-r^2$ are concerned. Naturally, 1111 behaves differently since it is a quasi-2D material. Another principal difference is the absence of the degeneracy in the corner of the BZ for 122 case.

In Figs. S6, S7 and S9 below we show the raw data from which the second derivatives were taken as well as the EDC curves from these raw data.

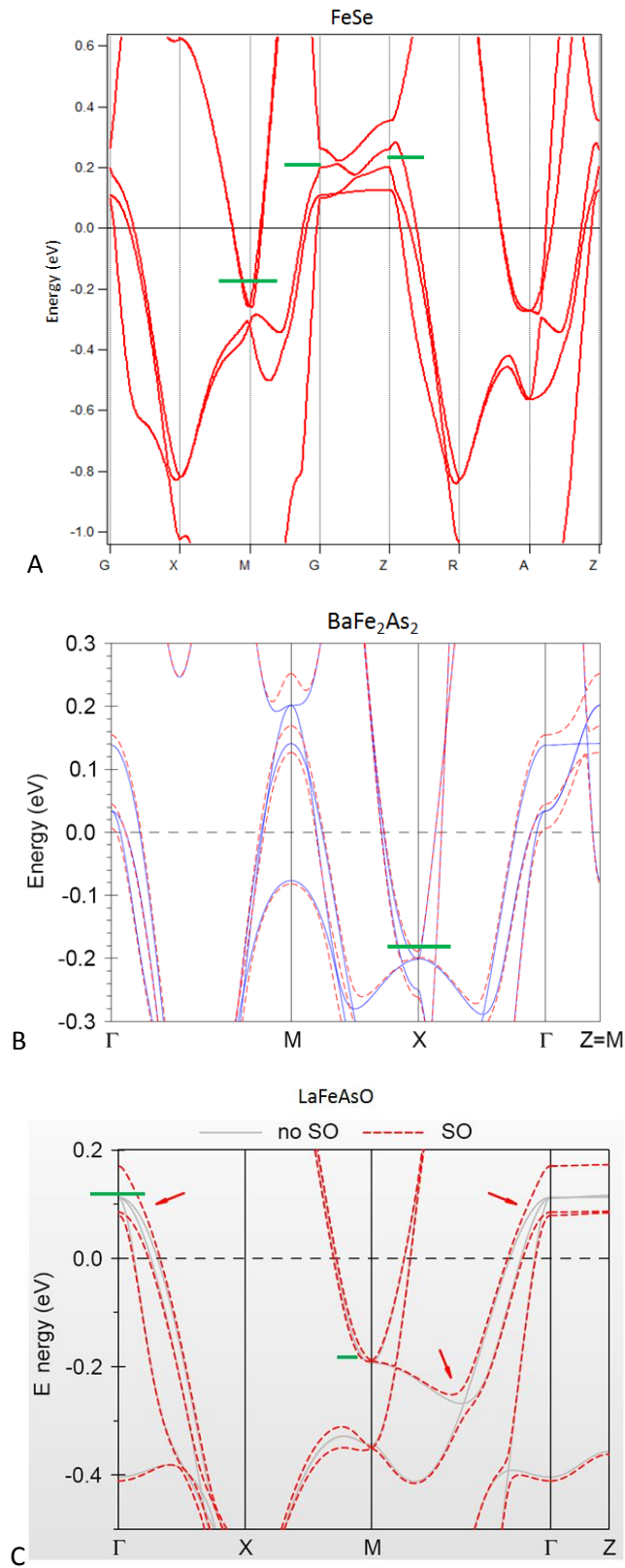

**Fig. S5.** Relativistic band-structure calculations of (A) FeSe, (B) BaFe$_2$As$_2$ and (C) LaFeAsO. Red curves – with SOC, blue or grey curves – without SOC. Green lines approximately indicate the position of the experimental Fermi level near Γ and M-points. Red arrows show where the SOC effect is the largest.

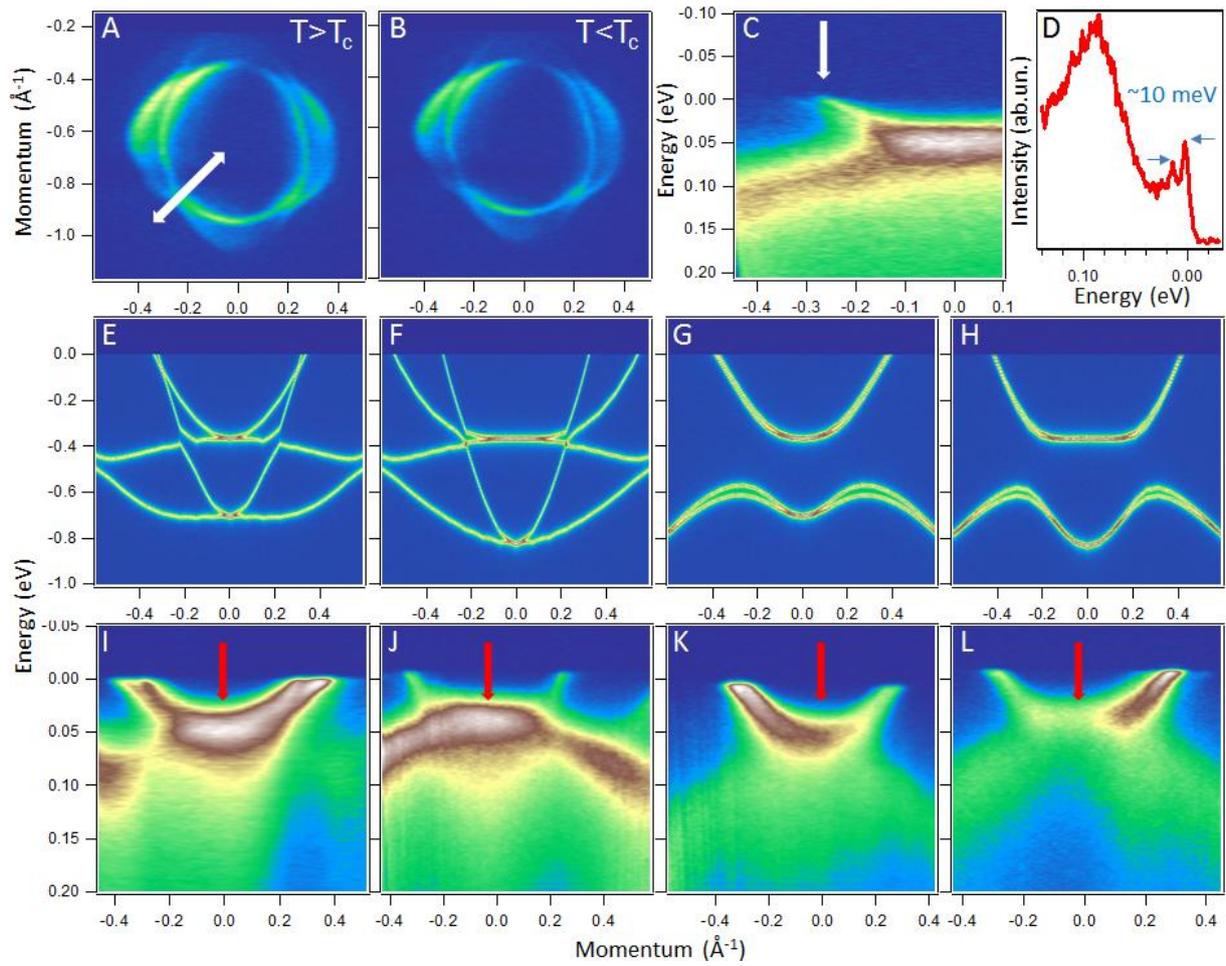

**Fig. S6.** The same as Fig. 3 of the main text, but with raw data instead of second derivatives.

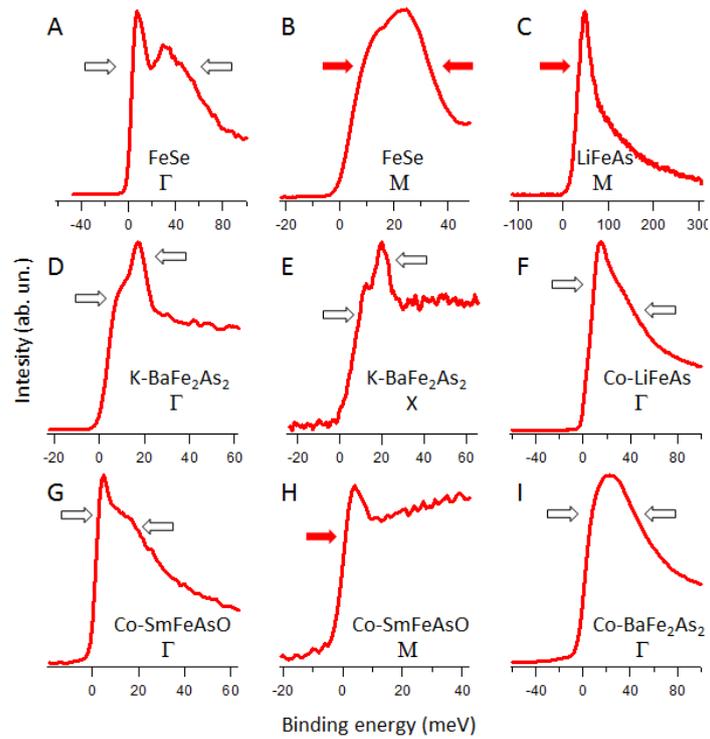

**Fig. S7** Energy distribution curves corresponding to zero momentum in cuts from Figs. 3 and 4 of the main text. (A) from Fig. 4 D, (B) from Fig. 4 F, (C) from Fig. 3 L, (D) from Fig. 4 H (E) from Fig. 4 I (F) EDC from Γ-point of 10% Co-LiFeAs sample, (G) from Fig. 4 J, (H) from Fig. 4 K(I) from Fig. 4 G.

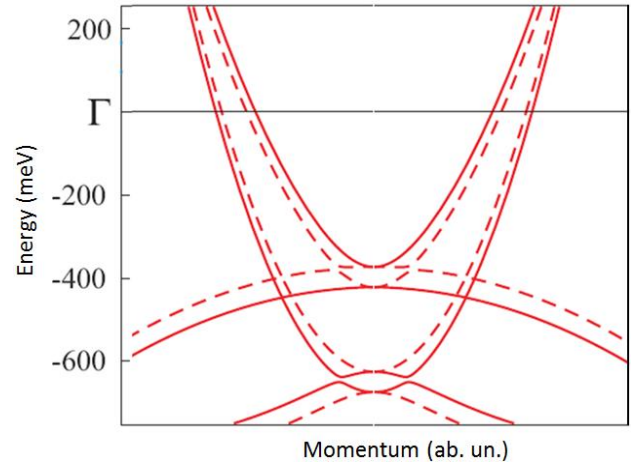

**Fig. S8.** Adapted from Fernandes and Vafek paper (Ref.21 main text). The cut is along the line connecting center and corner of the BZ and is centered on the latter. Solid and dashed lines correspond to two domains with mutually perpendicular nematic orderings. This superposition is supposed to be seen by ARPES.

In Fig. S8 we reproduce the result obtained recently by Fernandes and Vafek. They provided clear way to distinguish the nematic and SOC splittings. According to our data, we did not detect such effect in any of the studied materials.

**Fig. S9.** The same as Fig. 4 in the main text, but with the raw data instead of second derivatives.

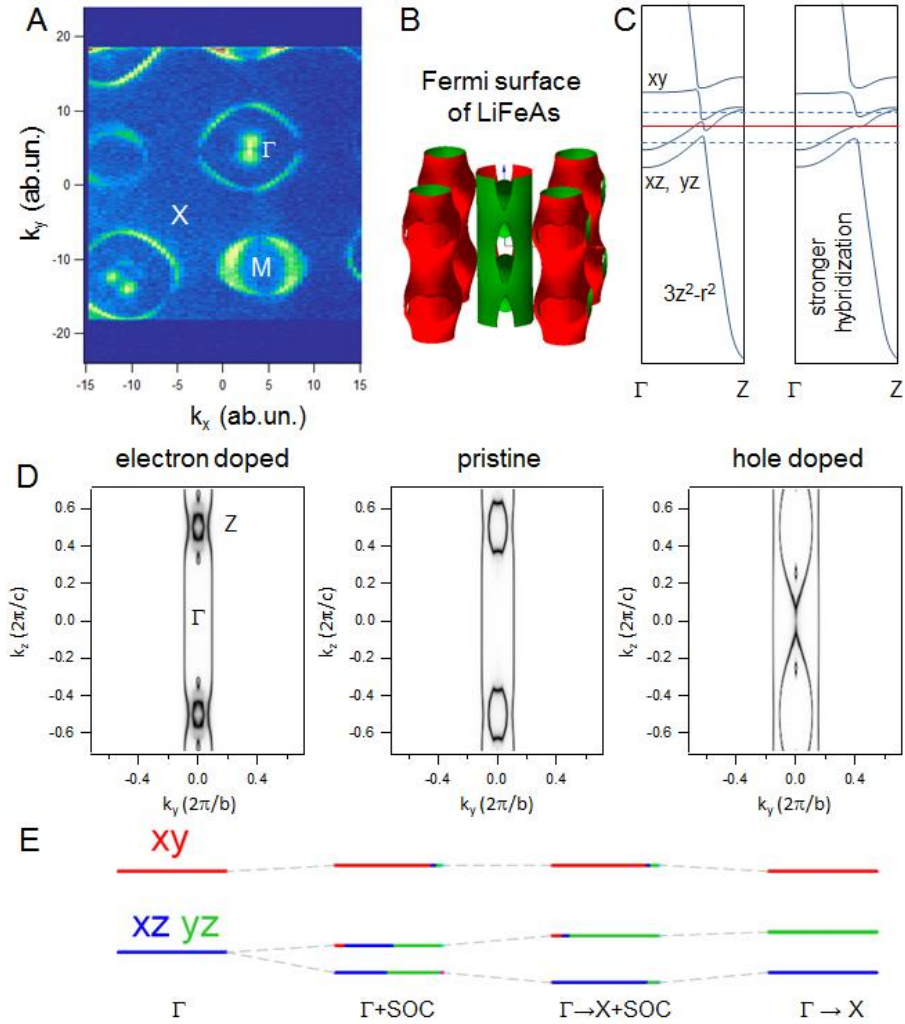

**Fig. S10.** (A) Experimental Fermi surface map taken using 80 eV photon energy. The signal is integrated within 10 meV window at the Fermi level. (B) Schematic representation of all three-dimensional Fermi surface sheets of LiFeAs. The openings in the larger Fermi surfaces are to visualize the presence of the smaller ones inside them. Red surfaces are electron-like. Green surfaces are hole-like. (C) Calculated band structure with SOC along ΓZ together with the sketch illustrating the stronger hybridization between $3z^2-r^2$ and other states. Red solid line – experimental Fermi level. Dashed lines position of the Fermi level upon hypothetical electron- and hole-doping (rigid shift). (D) Calculated Fermi surface section in $k_y$-$k_z$ plane for different doping levels (rigid shift). (E) Orbital mixing due to SOC near Γ and for ΓX-direction. The length of the color bar is proportional to the orbital weight. Small weights of $x^2-y^2$ and $3z^2-r^2$ caused by SOC are shown by magenta and cyan, respectively.

Conventional Fermi surface mapping, as shown in Fig. S10 A, allows one to identify the large Fermi surfaces around the center and around the corners of the BZ. The high-precision mapping (Fig. 3 A and B) revealed the separated inner and outer electron pockets, but the momentum distribution of the intensity right near Γ(Z) points is always complicated and depends on photon energy and geometry of the experiment (Fig. S10 A).

We use the information from Fig. 2 and the results of the calculations to reconstruct the full 3D Fermi surface of LiFeAs in Fig. S10 B. There are small three-dimensional hole-like pockets closed around Z-points. In the presence of spin-orbit coupling the orbital composition of the states supporting this Fermi surface is non-trivial as is seen from the sketches in Fig. S10 C: it depends on the degree of the hybridization between xz/yz and $3z^2-r^2$ states and their mixture because of SOC. We found that the Fermi level in stoichiometric LiFeAs runs through this complicated manyfold of bands as shown in Fig.

S10 C. We neglect the "wiggle" right at the Fermi level not only because our $k_z$ resolution is not sufficient to detect such features, but also because the feature itself can be flattened out by slightly larger hybridization, as shown in the right panel of Fig. S10 C. Nevertheless, with these new findings the Fermi surface topology of LiFeAs becomes very sensitive to the charge carriers concentration. Upon electron doping there could appear two additional electron-like small droplets above and below the Z-pockets (Fig. S10 D, left panel). Doping with holes will result in additional small hole pockets closer to $\Gamma$-point (Fig. S10 D, right panel).

[1] M. J. Pitcher et al. Chem. Comm. 101, 5918 (2008)

[2] A.Yu. Perlov and A.N. Yaresko and V.N. Antonov " *A Spin-polarized Relativistic Linear Muffin-tin Orbitals Package for Electronic Structure Calculations*", unpublished

[3] V. Antonov, B. Harmon and A. Yaresko in "*Electronic structure and magneto-optical properties of solids*", Kluwer Academic Publishers, Dordrecht, Boston, London, 2004

[4] I. Morozov et al. *Single Crystal Growth and Characterization of Superconducting LiFeAs.* Crystal Growth and Design, 10, 4428 (2010)